\documentclass[twoside,twocolumn,9pt]{article}
\usepackage{extsizes}
\usepackage[super,sort&compress,comma]{natbib} 
\usepackage[version=3]{mhchem}
\usepackage[left=1.5cm, right=1.5cm, top=1.785cm, bottom=2.0cm]{geometry}
\usepackage{balance}
\usepackage{mathptmx}
\usepackage{sectsty}
\usepackage{graphicx} 
\usepackage{lastpage}
\usepackage[format=plain,justification=justified,singlelinecheck=false,font={stretch=1.125,small,sf},labelfont=bf,labelsep=space]{caption}
\usepackage{float}
\usepackage{fancyhdr}
\usepackage{fnpos}
\usepackage{overpic}
\usepackage[english]{babel}
\addto{\captionsenglish}{%
  
}
\usepackage{array}
\usepackage{droidsans}
\usepackage{charter}
\usepackage[T1]{fontenc}
\usepackage[usenames,dvipsnames]{xcolor}
\usepackage{setspace}
\usepackage[compact]{titlesec}
\usepackage{hyperref}
\usepackage{bm}
\usepackage{amsmath,amssymb}

\usepackage{epstopdf}

\definecolor{cream}{RGB}{222,217,201}

\begin{document}

\pagestyle{fancy}
\thispagestyle{plain}
\fancypagestyle{plain}{
\renewcommand{\headrulewidth}{0pt}
}

\makeFNbottom
\makeatletter
\renewcommand\LARGE{\@setfontsize\LARGE{15pt}{17}}
\renewcommand\Large{\@setfontsize\Large{12pt}{14}}
\renewcommand\large{\@setfontsize\large{10pt}{12}}
\renewcommand\footnotesize{\@setfontsize\footnotesize{7pt}{10}}
\makeatother

\renewcommand{\thefootnote}{\fnsymbol{footnote}}
\renewcommand\footnoterule{\vspace*{1pt}%
\color{cream}\hrule width 3.5in height 0.4pt \color{black}\vspace*{5pt}} 
\setcounter{secnumdepth}{5}

\makeatletter 
\renewcommand\@biblabel[1]{#1}            
\renewcommand\@makefntext[1]%
{\noindent\makebox[0pt][r]{\@thefnmark\,}#1}
\makeatother 
\renewcommand{\figurename}{\small{Fig.}~}
\sectionfont{\sffamily\Large}
\subsectionfont{\normalsize}
\subsubsectionfont{\bf}
\setstretch{1.125} 
\setlength{\skip\footins}{0.8cm}
\setlength{\footnotesep}{0.25cm}
\setlength{\jot}{10pt}
\titlespacing*{\section}{0pt}{4pt}{4pt}
\titlespacing*{\subsection}{0pt}{15pt}{1pt}

\fancyfoot{}
\fancyfoot[LO,RE]{\vspace{-7.1pt}\includegraphics[height=9pt]{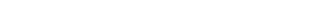}}
\fancyfoot[CO]{\vspace{-7.1pt}\hspace{13.2cm}\includegraphics{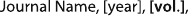}}
\fancyfoot[CE]{\vspace{-7.2pt}\hspace{-14.2cm}\includegraphics{head_foot/RF}}
\fancyfoot[RO]{\footnotesize{\sffamily{1--\pageref{LastPage} ~\textbar  \hspace{2pt}\thepage}}}
\fancyfoot[LE]{\footnotesize{\sffamily{\thepage~\textbar\hspace{3.45cm} 1--\pageref{LastPage}}}}
\fancyhead{}
\renewcommand{\headrulewidth}{0pt} 
\renewcommand{\footrulewidth}{0pt}
\setlength{\arrayrulewidth}{1pt}
\setlength{\columnsep}{6.5mm}
\setlength\bibsep{1pt}

\makeatletter 
\newlength{\figrulesep} 
\setlength{\figrulesep}{0.5\textfloatsep} 

\newcommand{\topfigrule}{\vspace*{-1pt}%
\noindent{\color{cream}\rule[-\figrulesep]{\columnwidth}{1.5pt}} }

\newcommand{\botfigrule}{\vspace*{-2pt}%
\noindent{\color{cream}\rule[\figrulesep]{\columnwidth}{1.5pt}} }

\newcommand{\dblfigrule}{\vspace*{-1pt}%
\noindent{\color{cream}\rule[-\figrulesep]{\textwidth}{1.5pt}} }

\makeatother

\twocolumn[
  \begin{@twocolumnfalse}
{\includegraphics[height=30pt]{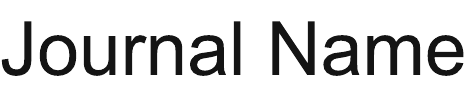}\hfill\raisebox{0pt}[0pt][0pt]{\includegraphics[height=55pt]{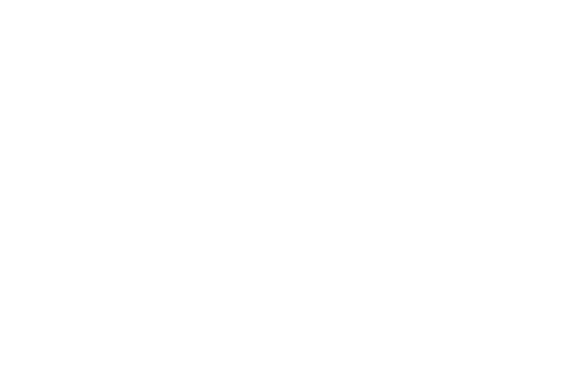}}\\[1ex]
\includegraphics[width=18.5cm]{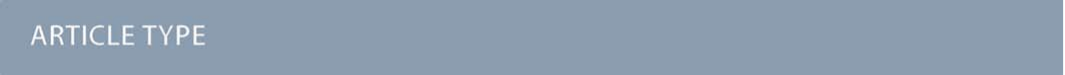}}\par
\vspace{1em}
\sffamily
\begin{tabular}{m{4.5cm} p{13.5cm} }

\includegraphics{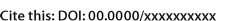} & \noindent\LARGE{\textbf{A unified description of flow-induced scission of wormlike micelles under shear and extensional flows}} \\

\vspace{0.3cm} & \vspace{0.3cm} \\

 & \noindent\large{Yusuke Koide,$^{\ast}$\textit{$^{a}$} Takato Ishida,\textit{$^{b}$} Takashi Uneyama,\textit{$^{b}$} and Yuichi Masubuchi\textit{$^{b}$}} \\

\includegraphics{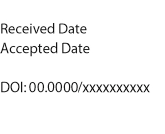} & \noindent\normalsize{
  We investigate flow-induced scission of wormlike micelles under different flow kinematics using dissipative particle dynamics simulations of surfactant solutions in shear, uniaxial extensional, planar extensional, and biaxial extensional flows.
  The average lifetime of wormlike micelles is used to quantify the degree of scission.
  While flow-induced scission occurs in all flow types when the deformation rate is sufficiently high, the dependence of the average lifetime on the deformation rate differs among flow types.
  To provide a unified description, we introduce an effective extension rate determined by the velocity gradient tensor and micellar orientation.
  When expressed in terms of this effective extension rate, the average lifetimes obtained under all flow types considered collapse onto a single curve.
  These results demonstrate that a unified description of flow-induced scission requires not only the strength and kinematics of the imposed flow, but also the micellar orientation relative to the extensional direction. 
  } \\

\end{tabular}

 \end{@twocolumnfalse} \vspace{0.6cm}

  ]

\renewcommand*\rmdefault{bch}\normalfont\upshape
\rmfamily
\section*{}
\vspace{-1cm}


\footnotetext{\textit{$^{a}$~Graduate School of Engineering Science, The University of Osaka, 1-3 Machikaneyama, Toyonaka, Osaka 560-8531, Japan. E-mail: y.koide.es@osaka-u.ac.jp}}
\footnotetext{\textit{$^{b}$~Department of Materials Physics, Graduate School of Engineering, Nagoya University,\\ Furo-cho, Chikusa, Nagoya, Aichi 464-8603, Japan}}





\section{Introduction}

Under suitable conditions, surfactants self-assemble into flexible, elongated aggregates known as wormlike micelles~\cite{Dreiss2007-jf}.
Although they are structurally similar to polymers, wormlike micelles undergo repeated scission and recombination. 
Micellar scission occurs due to thermal fluctuations at equilibrium and can be further promoted under strong flows.
This flow-induced scission leads to characteristic rheological properties and flow phenomena of wormlike micellar solutions.
For example, wormlike micellar solutions exhibit a nonmonotonic dependence of the steady-state extensional viscosity on the extension rate due to flow-induced scission of wormlike micelles~\cite{Prudhomme1994-tu,Walker1996-cf,Chen1997-xd,Koide2025-zr}.
Also, wormlike micelles can reduce friction drag in turbulent flows, and the critical flow strength above which the drag reduction begins to deteriorate has been linked to flow-induced scission of wormlike micelles in turbulence~\cite{Gyr1995-nn,Lu1998-au,Li2004-gt}.
Therefore, a fundamental understanding of flow-induced scission is crucial for predicting and controlling the fluid properties of wormlike micellar solutions.

Although micellar scission is difficult to observe directly in experiments, several studies have attempted to characterize the scission behavior of surfactant micelles using experimentally measurable quantities~\cite{Candau1990-he,Kern1992-tq,Couillet2004-mj}.
Candau \textit{et al.}~\cite{Candau1990-he} estimated the average breaking time of micelles in cetyltrimethylammonium bromide solutions by analyzing the relaxation of the light-scattering signal after a temperature jump~(T-jump), where a rapid and small change in temperature was imposed. 
By combining the T-jump results with rheological measurements, they discussed the physical mechanism underlying single-exponential stress relaxation.
Under flow, micellar deformation is expected to promote scission. 
Indeed, numerous experimental studies have reported indirect evidence suggesting the occurrence of flow-induced scission~\cite{Prudhomme1994-tu,Walker1996-cf,Rothstein2003-op,Chen2004-np,Arenas-Gomez2020-yy}.
For instance, Arenas-G\'{o}mez \textit{et al.}~\cite{Arenas-Gomez2020-yy} investigated the alignment of wormlike micelles under shear flow using small-angle neutron scattering~(SANS).
They reported a decrease in the orientation parameter at high shear rates, which may be attributed to a reduction in contour length caused by flow-induced scission.
Although experimental evidence for flow-induced micellar scission remains scarce, a few studies have reported observations consistent with its occurrence~\cite{Chen1997-xd,Huang2025-rs}.
For example, Huang \textit{et al.}~\cite{Huang2025-rs} analyzed SANS data using spherical harmonic decomposition and demonstrated that the average micellar length decreased at high shear rates under shear flow.

Owing to the difficulty of directly observing scission events of individual micelles in experiments, molecular simulations play an important complementary role.
Micellar scission properties have been investigated using several numerical approaches.
One approach is to evaluate the potential of mean force through an umbrella sampling method~\cite{Mandal2018-fs,Wang2018-db,Wand2020-qm,Mandal2022-il}.
Mandal \textit{et al.}~\cite{Mandal2018-fs} applied this method to cetyltrimethylammonium chloride micelles and evaluated the scission energy, which is the free energy difference associated with breaking a micelle into two shorter micelles.
By varying the salt ion ratio, they found a correlation between the obtained scission energy and the experimentally reported zero-shear viscosity.
Another approach is to apply a flow field to surfactant solutions, such as shear flow~\cite{Kroger1996-el,Padding2008-ap,Huang2009-nw,Sambasivam2015-xm,Koide2022-bp} and extensional flow~\cite{Dhakal2016-yk,Koide2025-zr}.
In our previous works~\cite{Koide2022-bp,Koide2025-zr}, we employed the dissipative particle dynamics~(DPD) method~\cite{Hoogerbrugge1992-ng,Espanol1995-mx} and evaluated the average lifetime of wormlike micelles under shear and uniaxial extensional flows.
We demonstrated that the average lifetime monotonically decreased with increasing deformation rate in both flows.
However, how the results obtained under different flows are related remains unclear, even though a unified understanding of micellar scission is important for developing coarse-grained models of wormlike micellar solutions~\cite{Vasquez2007-zd,Tamano2020-gd,Hommel2021-oi,Sato2022-rn}.

In the present study, we systematically investigate the flow-induced scission of wormlike micelles under various flow types to clarify the role of flow kinematics.
Specifically, we conduct DPD simulations of surfactant solutions under shear, uniaxial extensional, planar extensional, and biaxial extensional flows.
We evaluate the dependence of the average lifetime on the deformation rate under these flows.
A key feature of our approach is the introduction of an effective extension rate, defined from the velocity gradient tensor and the micellar orientation vector.
We demonstrate that the effective extension rate enables us to understand flow-induced scission under different flows in a unified manner.
\section{Method}
We employ the DPD method to study nonionic surfactant solutions.
Surfactants are modeled as coarse-grained chains consisting of one hydrophilic head particle and two hydrophobic tail particles~\cite{Li2019-wy,Koide2022-bp,Koide2025-zr}.
Adjacent particles along a surfactant molecule are connected by a harmonic bond force 
\begin{equation}
    \bm{F}_{ij}^\mathrm{B} = -k_\mathrm{s} (|\bm{r}_{ij}|-r_\mathrm{eq}) \bm{e}_{ij}, \label{eq:bond_force} 
\end{equation}
where $k_\mathrm{s}$ is the spring constant, $r_\mathrm{eq}$ is the equilibrium bond distance, $\bm{r}_{ij}=\bm{r}_i-\bm{r}_j$, and $\bm{e}_{ij}=\bm{r}_{ij}/|\bm{r}_{ij}|$ with $\bm{r}_i$ being the position of the $i$-th particle.
Water particles are also included in the system.
These DPD particles, including surfactant and water particles, interact through repulsive, dissipative, and random forces.
The dissipative force is evaluated using the difference between the particle velocities in the laboratory frame.
Further details of the DPD model, such as the functional forms of the interaction forces, can be found in previous publications~\cite{Koide2022-bp,Koide2023-ao}.
In the following, all quantities are nondimensionalized by $k_\mathrm{B}T_0$, $m$, and $r_\mathrm{c}$, where $k_\mathrm{B}$ is the Boltzmann constant, $T_0$ is the reference temperature, $m$ is the mass of a DPD particle, and $r_\mathrm{c}$ is the cutoff distance.

In the present study, the simulation parameters are set as follows:
the total number of particles is $N=648\,000$; the number density of particles is $\rho=3$; the dissipative force coefficient is $\gamma=4.5$; the volume fraction of surfactant particles is $\phi=0.05$; the spring constant is $k_\mathrm{s}=50$; the equilibrium bond length is $r_\mathrm{eq}=0.8$; the repulsive force coefficients are $a_\mathrm{hh}=25$, $a_\mathrm{ht}=60$, $a_\mathrm{hw}=20$, $a_\mathrm{tt}=25$, $a_\mathrm{tw}=60$, and $a_\mathrm{ww}=25$, where h, t, and w denote head, tail, and water particles, respectively.
These values of $a_\mathrm{ij}$ are taken from previous studies~\cite{Li2019-wy,Koide2022-bp,Koide2025-zr}, where surfactants form wormlike micelles above a certain volume fraction $\phi$.
At $\phi=0.05$, a sufficient number of wormlike micelles are available for statistical analysis.
Although the soft-core potential employed in the DPD method cannot capture entanglement effects, as noted previously for polymer systems~\cite{Pan2002-oq}, this is not a concern in the present study, which focuses on unentangled wormlike micelles.
We systematically investigate flow-induced scission of wormlike micelles by varying the temperature $T$.
When $T$ is varied, the random force coefficient $\sigma$ is adjusted to satisfy the fluctuation-dissipation relation $\sigma^2=2\gamma k_\mathrm{B}T$ while keeping $\gamma$ fixed at $4.5$.

\begin{figure}
  \centering
  \begin{overpic}[width=0.65\linewidth]{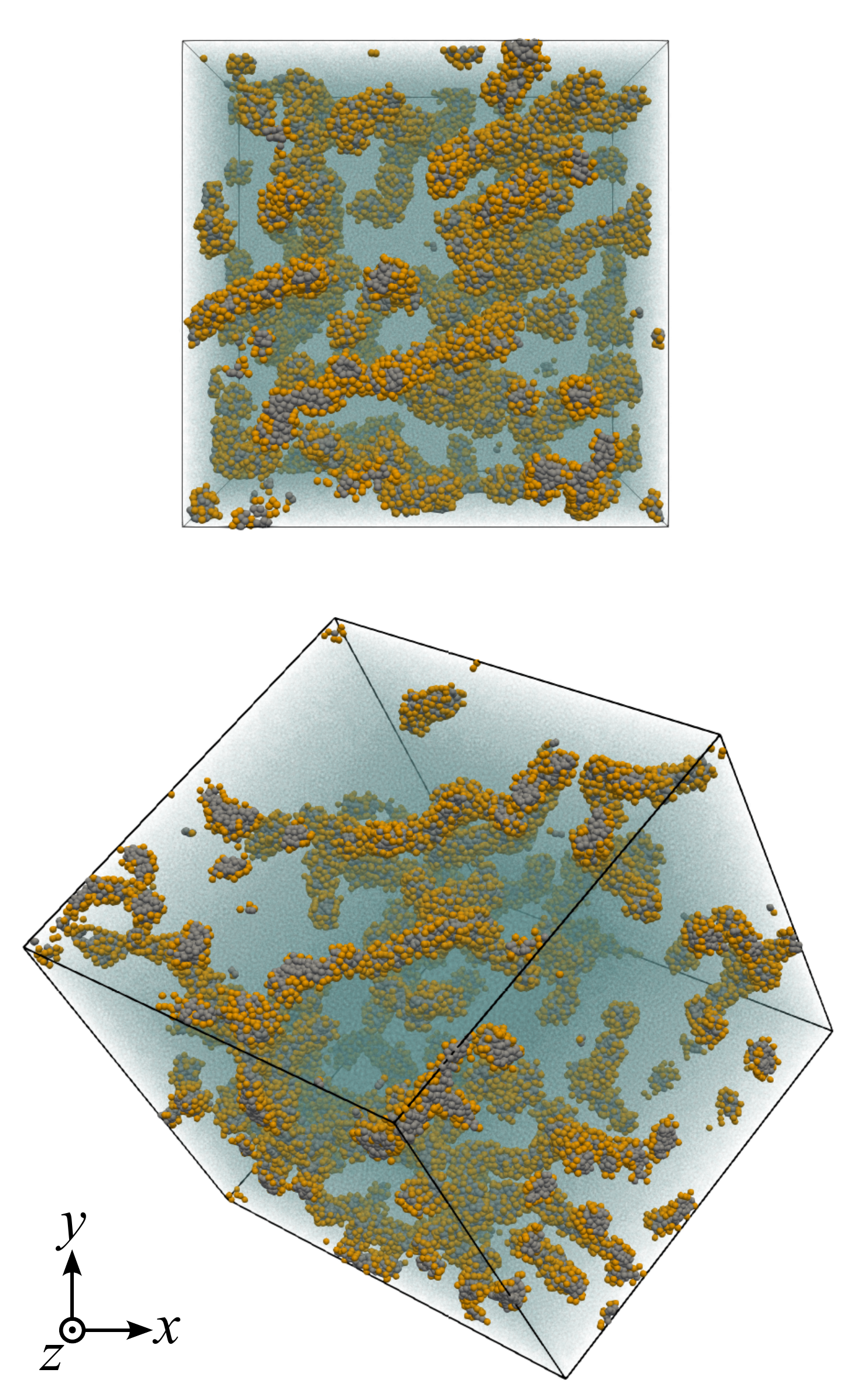} 
      \put(6,95){(a)}
      \put(6,50){(b)}
  \end{overpic}
  \caption{Snapshots of the surfactant solution under (a) shear flow at $\dot{\gamma}=0.002$ and (b) biaxial extensional flow at $\dot{\epsilon}_\mathrm{B}=0.002$, with $k_\mathrm{B}T=1$. Hydrophilic and hydrophobic particles are shown in orange and gray, respectively. For clarity, water particles are represented by semitransparent blue dots.}
  \label{fig:snapshot}
\end{figure}

To reveal the effect of flow kinematics on flow-induced scission of wormlike micelles, we consider four flow types: shear, uniaxial extensional, planar extensional, and biaxial extensional flows.
The particle motion is governed by SLLOD equations~\cite{evans_morriss_2008}
\begin{align}
  &\frac{d{\bm{r}_i}}{dt} = {\bm{p}_i} + (\nabla \bm{u})^\mathsf{T} \cdot\bm{r}_i \\
  &\frac{d{\bm{p}}_i}{dt} = \bm{F}_i - (\nabla \bm{u})^\mathsf{T} \cdot\bm{p}_i ,
\label{eq:SLLOD_equation}
\end{align}
where $\bm{p}_i$ is the peculiar momentum of the $i$-th particle, $\bm{F}_i$ is the force acting on the $i$-th particle, including the repulsive, dissipative, random, and bond forces, and $\nabla\bm{u}$ is the velocity gradient tensor of the imposed flow field.
For shear, uniaxial extensional, planar extensional, and biaxial extensional flows, $\nabla\bm{u}$ is given by
\begin{align}
  \left(\nabla\bm{u}\right)_\mathrm{S} &=
  \begin{pmatrix}
      0 & 0 & 0\\
      \dot{\gamma} & 0 & 0\\
      0 & 0 & 0
  \end{pmatrix},
  \label{eq:shear}\\
  \left(\nabla\bm{u}\right)_\mathrm{E} &=
  \begin{pmatrix}
      \dot{\epsilon} & 0 & 0\\
      0 & -\dot{\epsilon}/2 & 0\\
      0 & 0 & -\dot{\epsilon}/2
  \end{pmatrix},
  \label{eq:uniaxial}\\
  \left(\nabla\bm{u}\right)_\mathrm{P} &=
  \begin{pmatrix}
      \dot{\epsilon}_\mathrm{P} & 0 & 0\\
      0 & -\dot{\epsilon}_\mathrm{P} & 0\\
      0 & 0 & 0
  \end{pmatrix},
  \label{eq:planar}\\
  \left(\nabla\bm{u}\right)_\mathrm{B} &=
  \begin{pmatrix}
      \dot{\epsilon}_\mathrm{B} & 0 & 0\\
      0 & \dot{\epsilon}_\mathrm{B} & 0\\
      0 & 0 & -2\dot{\epsilon}_\mathrm{B}
  \end{pmatrix}.
  \label{eq:biaxial}
\end{align}
Here, $\dot{\gamma}$ is the shear rate, $\dot{\epsilon}$ is the extension rate, $\dot{\epsilon}_\mathrm{P}$ is the planar extension rate, and $\dot{\epsilon}_\mathrm{B}$ is the biaxial extension rate. 
Figure~\ref{fig:snapshot} shows representative snapshots of surfactant solutions under shear and biaxial extensional flows.
For shear flow, we use the Lees-Edwards boundary condition~\cite{Lees1972-mx}.
For uniaxial, planar, and biaxial extensional flows, we use the generalized Kraynik--Reinelt (GKR) boundary conditions~\cite{Dobson2014-kr,Hunt2016-vl}, where the simulation box deforms according to the imposed flow and is systematically remapped to prevent excessive deformation.

We now describe the details of the numerical implementation.
The modified velocity Verlet method~\cite{Groot1997-je} is used for time integration, where the parameter $\lambda$ in this scheme and the time step $\Delta t$ are set to $0.65$ and $0.04$, respectively.
We have confirmed that these parameters maintain the temperature within a relative error of $1\,\%$, even under the strong flows with large velocity gradients considered here.
The initial simulation box is a cube with dimensions $60\times 60\times 60$.
We conduct equilibrium simulations for $20\,000$ time units from a random initial configuration until the potential energy and the number of micelles reach statistically steady values.
After this initial equilibration, we conduct nonequilibrium simulations.
Since the present study focuses on scission behavior in a statistically steady state, all analyses are performed $4\,000$ time units after the onset of flow.
Because the longest relaxation time of wormlike micelles is below $1\,500$ time units in all cases, both the initial equilibration period and the waiting time after the onset of flow are sufficiently long.
All DPD simulations are conducted with our in-house code.
\section{Results and Discussion}

\begin{figure*}
  \centering
  \begin{overpic}[width=1\linewidth]{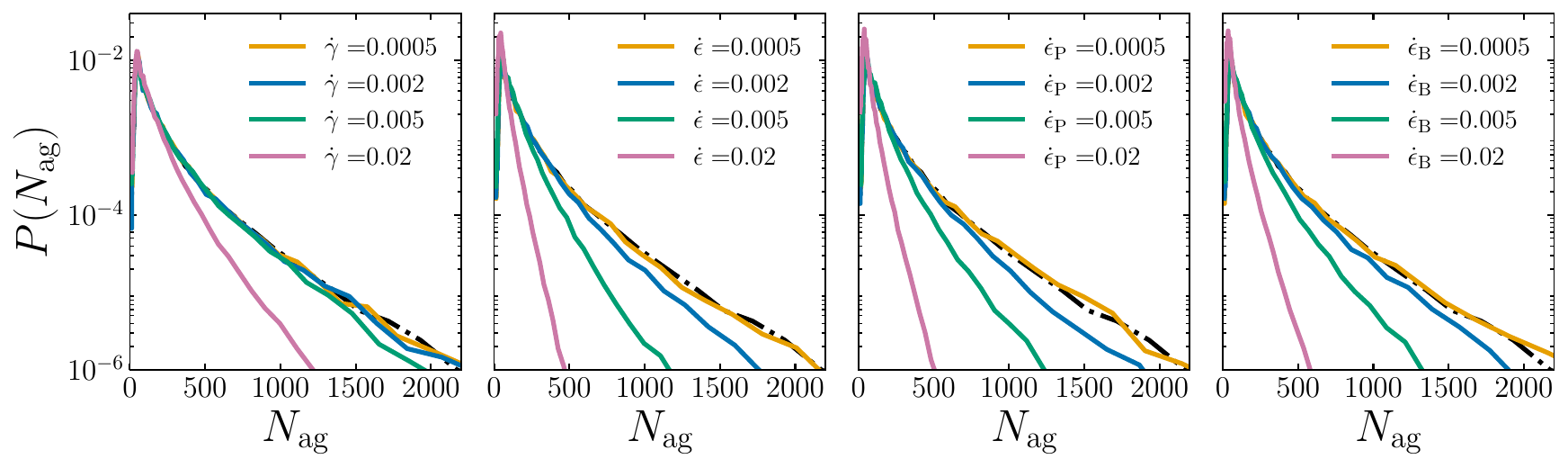} 
            \put(8,30){(a)}
            \put(31.5,30){(b)}
            \put(54.5,30){(c)}
            \put(77.5,30){(d)}
  \end{overpic}
  \caption{Probability density function $P(N_\mathrm{ag})$ of the aggregation number $N_\mathrm{ag}$ at $k_\mathrm{B}T=1$ for (a) shear flow, (b) uniaxial extensional flow, (c) planar extensional flow, and (d) biaxial extensional flow. The black dash-dotted line represents the result at equilibrium.}

  \label{fig:nag_pdf}
\end{figure*}
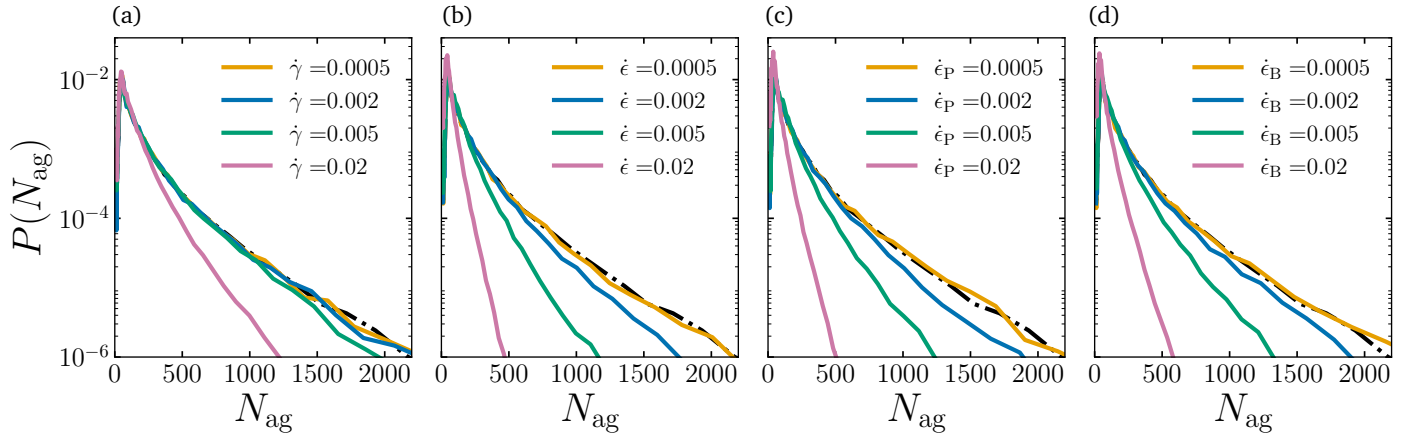

To demonstrate the occurrence of flow-induced scission of surfactant micelles, we first investigate the aggregation number distribution of micelles.
We define a micelle using the same method as in previous studies.~\cite{Vishnyakov2013-ge,Lee2016-cx}
In this method, two surfactant molecules belong to the same cluster if a hydrophobic particle of one molecule is within $r_c(=1)$ of a hydrophobic particle of the other.
If a cluster has an aggregation number $N_\mathrm{ag}$ greater than a threshold value $n_\mathrm{mic}$, the cluster is regarded as a micelle.
We set $n_\mathrm{mic}$ to $10$ as in previous studies~\cite{Koide2022-bp,Koide2025-zr}, and we have verified that variations in $n_\mathrm{mic}$ have negligible effects on the following results within the range $5\leq n_\mathrm{mic} \leq 20$. 
Figure~\ref{fig:nag_pdf} shows the probability density function~(PDF) $P(N_\mathrm{ag})$ of $N_\mathrm{ag}$ for each flow type at various deformation rates.
Unless otherwise noted, we present the results at $k_\mathrm{B}T=1$.
In our system, micelles are spherical for $N_\mathrm{ag}\lesssim 50$, rodlike for $50\lesssim N_\mathrm{ag}\lesssim 200$, and wormlike for $N_\mathrm{ag}\gtrsim 200$~\cite{Koide2022-bp}.
At the weakest deformation rate, $P(N_\mathrm{ag})$ almost overlaps with that in equilibrium.
As the deformation rate increases, $P(N_\mathrm{ag})$ gradually decreases at large $N_\mathrm{ag}$ regardless of the flow type, indicating that flow-induced scission of micelles occurs. 
Quantitatively, for fixed values of $\dot{\gamma}$, $\dot{\epsilon}$, $\dot{\epsilon}_\mathrm{P}$, and $\dot{\epsilon}_\mathrm{B}$~($\geq 0.002$), $P(N_\mathrm{ag})$ exhibits a more significant decrease at large $N_\mathrm{ag}$ for extensional flows than for shear flows.

To characterize the scission behavior of individual surfactant micelles, we focus on the micellar lifetime $t_b$, defined as the time interval between the formation and scission of a micelle.
Here, scission is defined as an event in which $N_\mathrm{ag}$ of a micelle decreases by more than the threshold value $n_\mathrm{mic}$ within a given time interval $\delta t(=100\Delta t)$~\cite{Koide2022-bp,Koide2023-ao}.
The value of $\delta t$ is chosen to be sufficiently short compared with the characteristic timescales of micellar scission, except at very large deformation rates or for very large $N_\mathrm{ag}$.
We evaluate the statistical properties of $t_b$ for each $N_\mathrm{ag}$ using the survival function $S(t)=\Pr (t_b>t)$, which represents the probability that micelles with $N_\mathrm{ag}$ survive beyond a certain time $t$.
Using the Kaplan--Meier method~\cite{Kaplan1958-yn}, we estimate $S(t)$ from lifetimes of micelles whose aggregation numbers lie within the range $[N_\mathrm{ag}-\Delta N_\mathrm{ag}/2,N_\mathrm{ag}+\Delta N_\mathrm{ag}/2]$, where $\Delta N_\mathrm{ag}$ is set to $0.1N_\mathrm{ag}$.
Specifically, given the obtained lifetimes $t_b^{(1)}<t_b^{(2)}<\ldots<t_b^{(n)}$, $S(t)$ is calculated as 
\begin{equation}
    S(t)= \prod_{\{j:t_b^{(j)}\leq t\}}\left(1-\frac{d_j}{r_j}\right) \label{eq:kaplan_meier},
\end{equation}
where $d_j$ is the number of scission events observed at $t_b^{(j)}$ and $r_j$ is the number of micelles surviving just before $t_b^{(j)}$.
In the Kaplan--Meier method, $r_j$ includes not only micelles that eventually undergo scission but also those that eventually undergo recombination.  
Note that the present study analyzes $t_b$ measured after the system has reached a statistically steady state following the onset of flow. 
In this state, micellar scission and recombination occur repeatedly with stationary statistics.

Using $S(t)$, we examine how the flow type and deformation rate affect flow-induced scission of individual micelles.
In what follows, to focus on flow effects, we present results for the fixed aggregation number $N_\mathrm{ag}=300$, which is chosen as a representative wormlike micelle.
Our conclusions remain essentially unchanged for $N_\mathrm{ag}\gtrsim 200$~(see Appendix A).
Figure~\ref{fig:survival_function} shows $S(t)$ for $N_\mathrm{ag}=300$ at the same values of $\dot{\gamma}$, $\dot{\epsilon}$, $\dot{\epsilon}_\mathrm{P}$, and $\dot{\epsilon}_\mathrm{B}$ as in Fig.~\ref{fig:nag_pdf}.
To show the statistical accuracy, we also present the 95\% confidence intervals obtained using the exponential Greenwood formula~\cite{kalbfleisch2011statistical}.
We observe that $S(t)$ exhibits an exponential decay irrespective of the flow type and deformation rate, as already reported for shear flow\cite{Koide2022-bp} and uniaxial extensional flow~\cite{Koide2025-zr}.
This exponential decay indicates that the probability of micellar scission per unit time is independent of elapsed time.
Quantitatively, the decay rate of $S(t)$ depends on the flow type and deformation rate.
At the lowest deformation rate shown in Fig.~\ref{fig:survival_function}, $S(t)$ almost coincides with that in equilibrium.
Thus, weak flows hardly affect micellar scission.
This tendency is consistent with the observation for $P(N_\mathrm{ag})$~(Fig.~\ref{fig:nag_pdf}).
As the deformation rate increases, $S(t)$ decays faster.
These results provide evidence that flow-induced scission of micelles with $N_\mathrm{ag}=300$ occurs under sufficiently strong flows irrespective of the flow type.
However, the degree of flow-induced scission~(i.e., the decay rate of $S(t)$) significantly depends on the flow type when comparing $S(t)$ at fixed values of $\dot{\gamma}$, $\dot{\epsilon}$, $\dot{\epsilon}_\mathrm{P}$, and $\dot{\epsilon}_\mathrm{B}$.
In particular, extensional flows promote micellar scission more than shear flows.
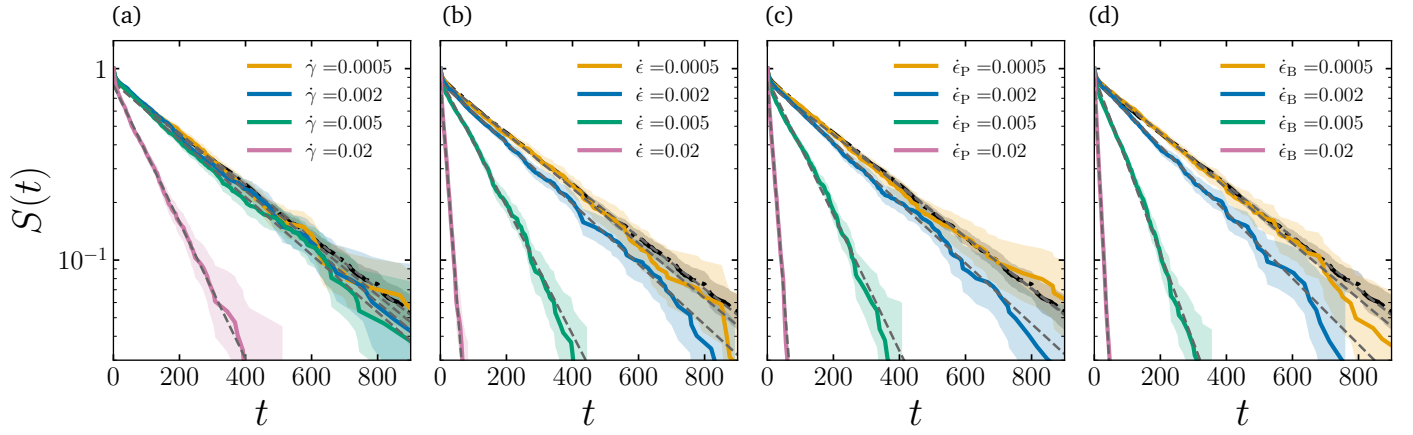
\begin{figure*}
  \centering
  \begin{overpic}[width=1\linewidth]{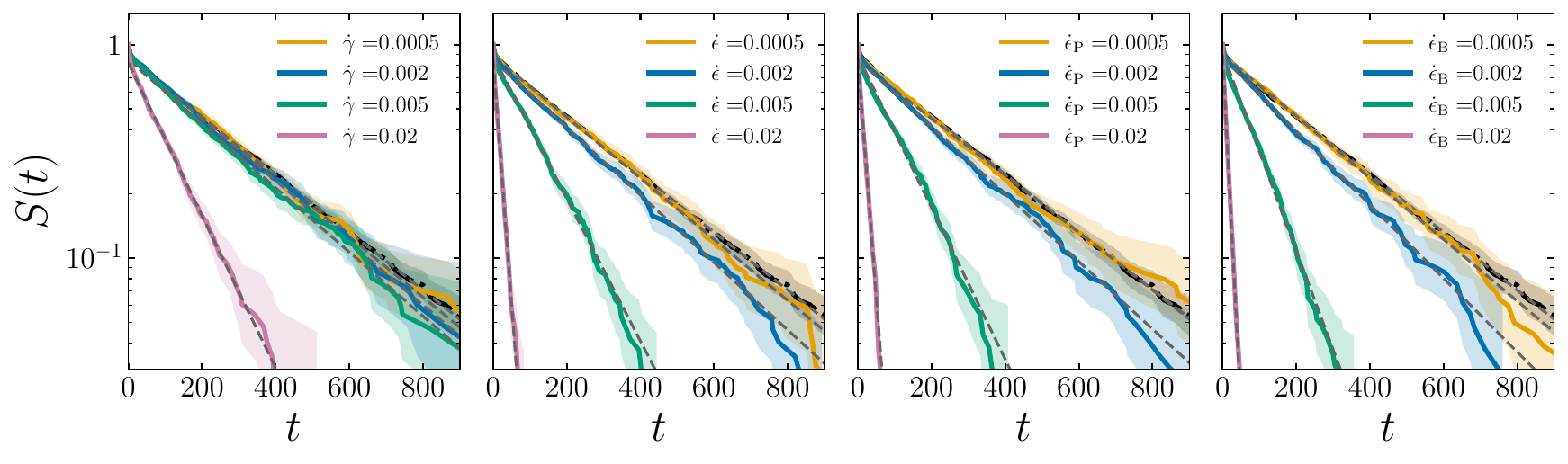} 
            \put(8,30){(a)}
            \put(31.5,30){(b)}
            \put(54.5,30){(c)}
            \put(77.5,30){(d)}
  \end{overpic}
  \caption{Survival function $S(t)$ of micelles with $N_\mathrm{ag}=300$ at $k_\mathrm{B}T=1$ for (a) shear flow, (b) uniaxial extensional flow, (c) planar extensional flow, and (d) biaxial extensional flow. The shaded regions indicate the $95\%$ confidence interval. The gray dashed lines indicate exponential fits to $S(t)$. The black dash-dotted line represents the result at equilibrium.}
  \label{fig:survival_function}
\end{figure*}%

\begin{figure}
    \centering
        \begin{overpic}[width=0.8\linewidth]{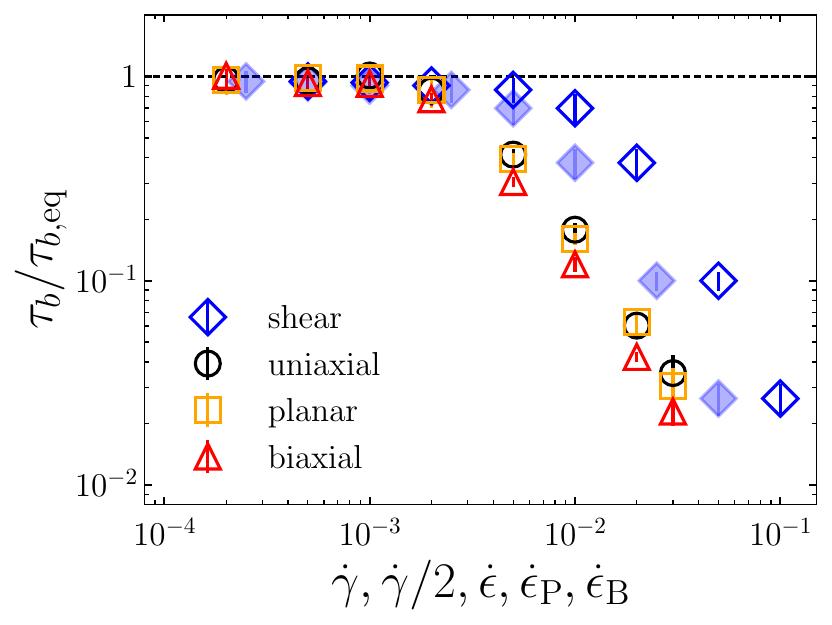}
            \linethickness{3pt}
        \end{overpic}
  
        \caption{Normalized average lifetime $\tau_b/\tau_{b,\mathrm{eq}}$ of micelles with $N_\mathrm{ag}=300$ at $k_\mathrm{B}T=1$ as a function of $\dot{\gamma}$, $\dot{\epsilon}$, $\dot{\epsilon}_\mathrm{P}$, and $\dot{\epsilon}_\mathrm{B}$ for shear flow, uniaxial extensional flow, planar extensional flow, and biaxial extensional flow, respectively. The semitransparent blue diamonds represent the shear-flow data replotted as a function of $\dot{\gamma}/2$. The black dashed line indicates $\tau_b/\tau_{b,\mathrm{eq}}=1$. The error bars denote the standard deviations from six independent simulations.}
        \label{fig:lifetime_flow_type}
  \end{figure}

To systematically investigate how micellar scission depends on flow type and deformation rate, we focus on the average lifetime $\tau_b$ of micelles for each $N_\mathrm{ag}$.
We obtain $\tau_b$ by fitting $S(t)$ to an exponential function of the form $S(t)\propto\exp(-t/\tau_b)$, as shown in Fig.~\ref{fig:survival_function}.
Figure~\ref{fig:lifetime_flow_type} shows $\tau_b$ for micelles with $N_\mathrm{ag}=300$ as a function of $\dot{\gamma}$, $\dot{\epsilon}$, $\dot{\epsilon}_\mathrm{P}$, and $\dot{\epsilon}_\mathrm{B}$.
Here, $\tau_b$ is normalized by its equilibrium value $\tau_{b,\mathrm{eq}}$.
For each flow type, $\tau_b/\tau_{b,\mathrm{eq}}\approx 1$ at low deformation rates, whereas $\tau_b/\tau_{b,\mathrm{eq}}$ becomes smaller than unity at sufficiently high deformation rates.
The decrease in $\tau_b/\tau_{b,\mathrm{eq}}$ demonstrates that flow-induced scission occurs under strong flows, as previously reported for shear and uniaxial extensional flows~\cite{Koide2022-bp,Koide2025-zr}.
Regarding the flow-type dependence, extensional flows promote micellar scission more strongly than shear flows, as already seen in Fig.~\ref{fig:survival_function}.
In addition, $\tau_b/\tau_{b,\mathrm{eq}}$ exhibits similar behavior under uniaxial and planar extensional flows, whereas it is slightly smaller under biaxial extensional flow.
A central aim of the present study is therefore to establish a unified description of $\tau_b$ across different flow types.

One may argue that a direct comparison based on $\dot{\gamma}$, $\dot{\epsilon}$, $\dot{\epsilon}_\mathrm{P}$, and $\dot{\epsilon}_\mathrm{B}$ is not appropriate because $\dot{\gamma}$ contains both extensional and rotational components.
To isolate the purely extensional component, which is likely to be more relevant to flow-induced scission, we focus on the strain-rate tensor $\bm{D}=[(\nabla \bm{u})+(\nabla \bm{u})^\mathsf{T}]/2$ and, in particular, on its largest eigenvalue. This eigenvalue is $\dot{\gamma}/2$, $\dot{\epsilon}$, $\dot{\epsilon}_\mathrm{P}$, and $\dot{\epsilon}_\mathrm{B}$ for shear, uniaxial extensional, planar extensional, and biaxial extensional flows, respectively.
In Fig.~\ref{fig:lifetime_flow_type}, we also plot $\tau_b/\tau_{b,\mathrm{eq}}$ for micelles with $N_\mathrm{ag}=300$ under shear flow as a function of $\dot{\gamma}/2$.
Although the difference in $\tau_b/\tau_{b,\mathrm{eq}}$ between shear flow and extensional flows is reduced when the purely extensional component is considered, this still cannot fully explain the more efficient flow-induced scission under extensional flows.
In addition, the reason why $\tau_b/\tau_{b,\mathrm{eq}}$ decreases slightly more rapidly in biaxial extensional flow than in uniaxial and planar extensional flows remains unclear.
This difference cannot be accounted for by the largest eigenvalue of the strain-rate tensor, which is identical by definition for the extensional flows considered here.

More generally, various kinematic measures and flow-classification schemes have been proposed in the field of rheology.~\cite{Astarita1979-xm,Fuller1980-ag,Thompson2005-jl,Poole2023-gk}
Although an exhaustive comparison of such measures is beyond the scope of the present study, we also examined $\sqrt{2\bm{D}:\bm{D}}$ as a representative invariant measure of the deformation rate; for example, this quantity reduces to $\dot{\gamma}$ in shear flow.
However, we confirmed that this measure did not improve the collapse of $\tau_b/\tau_{b,\mathrm{eq}}$ among the different flow types (data not shown).
Both the largest eigenvalue of $\bm{D}$ and $\sqrt{2\bm{D}:\bm{D}}$ are determined solely by the imposed velocity gradient tensor and contain no information about the micellar state.
Thus, the remaining differences in $\tau_b/\tau_{b,\mathrm{eq}}$ among the different flow types~(Fig.~\ref{fig:lifetime_flow_type}) indicate that micellar conformation also plays an essential role in flow-induced scission.

We then aim to describe the deformation-rate dependence of $\tau_b$ for different flow types in a unified way by introducing a relevant quantity that incorporates the effects of both flow kinematics and micellar conformation.
Specifically, we focus on the effective extension rate $\dot{\epsilon}_\mathrm{eff}$, defined as 
\begin{equation}
  \dot{\epsilon}_\mathrm{eff} = \left\langle \hat{\bm{\ell}}\cdot \left(\nabla \bm{u}\right)\cdot \hat{\bm{\ell}}\right \rangle_{N_\mathrm{ag}}= \left\langle \hat{\bm{\ell}}\cdot \bm{D}\cdot \hat{\bm{\ell}}\right \rangle_{N_\mathrm{ag}}.\label{eq:eps_eff}
\end{equation}
Here, $\hat{\bm{\ell}}$ denotes the instantaneous unit direction vector of a micelle, defined as the normalized eigenvector of the instantaneous gyration tensor $\bm{G}$ corresponding to its largest eigenvalue.
The tensor $\bm{G}$ is given by 
\begin{equation}
  \bm{G} = \frac{1}{N_\mathrm{sur}} \sum_{i=1}^{N_\mathrm{sur}}(\bm{r}_i-\bm{r}_G)(\bm{r}_i-\bm{r}_G),
\end{equation}
where $N_\mathrm{sur}(=3N_\mathrm{ag})$ is the number of surfactant particles in the micelle, $\bm{r}_i$ is the position of the $i$-th particle, and $\bm{r}_G$ is the position of the center of mass of the micelle.
Because $\hat{\bm{\ell}}$ varies among micelles and fluctuates in time, we characterize the average extension experienced by the micelles using the statistical average $\langle \cdot \rangle_{N_\mathrm{ag}}$, taken over micelles whose aggregation numbers lie within $[N_\mathrm{ag}-\Delta N_\mathrm{ag}/2,N_\mathrm{ag}+\Delta N_\mathrm{ag}/2]$.
Hereafter, the subscript $N_\mathrm{ag}$ is omitted for brevity.
For an individual micelle at a given instant, $\hat{\bm{\ell}}\cdot(\nabla\bm{u})\cdot\hat{\bm{\ell}}$ is the \(\hat{\bm{\ell}}\)-component of the directional derivative of the imposed velocity field in the \(\hat{\bm{\ell}}\) direction.
Thus, $\dot{\epsilon}_\mathrm{eff}$ represents the average rate of extension or compression along the micellar direction.
In addition, $\dot{\epsilon}_\mathrm{eff}$ is invariant under rotation of the coordinate system.
The corresponding quantity has been employed to analyze the stretching of material lines and dumbbell models in turbulence.~\cite{Girimaji1990-fs,Kida2002-gv,Watanabe2010-oi,Koide2024-lc}
A derivation of $\dot{\epsilon}_\mathrm{eff}$ for material lines is provided in Appendix B.

It is worth noting that a conceptually similar effective extension rate was recently introduced by King and McKinley to relate the shear viscosity to the planar extensional viscosity.~\cite{Nicholas2026-pt}
They projected $\bm{D}$ onto the principal direction of the stress tensor and defined $\dot{\epsilon}_\mathrm{eff}^\sigma=\bm{e}_\sigma\cdot\bm{D}\cdot\bm{e}_\sigma$, where $\bm{e}_\sigma$ denotes the eigenvector of the stress tensor corresponding to its largest eigenvalue. 
Although the present definition of $\dot{\epsilon}_\mathrm{eff}$~[Eq.~\eqref{eq:eps_eff}] is mathematically analogous to that of $\dot{\epsilon}_\mathrm{eff}^\sigma$, we use the instantaneous micellar direction $\hat{\bm{\ell}}$ instead of the principal direction of the stress tensor.
This difference in the projection direction reflects the different response of interest: their formulation characterizes a macroscopic rheological response, whereas the present formulation characterizes the stretching rate experienced along individual micelles.

To clarify the physical meaning of $\dot{\epsilon}_\mathrm{eff}$, it is useful to express $\dot{\epsilon}_\mathrm{eff}$ in terms of the principal directions of the strain-rate tensor.
Using the eigenvalues $\lambda_i~(\lambda_1\geq \lambda_2\geq \lambda_3)$ of $\bm{D}$, $\dot{\epsilon}_\mathrm{eff}$ can be written as 
\begin{equation}
  \dot{\epsilon}_\mathrm{eff} = \sum_{i=1}^3\lambda_i\langle\cos^2\theta_i\rangle, \label{eq:eps_eff_eigen}
\end{equation}
where $\theta_i$ is the angle between $\hat{\bm{\ell}}$ and the eigenvector $\bm{e}_i$ of $\bm{D}$.
For incompressible fluids, the eigenvalues satisfy $\lambda_1+\lambda_2+\lambda_3=0$ as a consequence of $\nabla\cdot\bm{u}=0$, which implies $\lambda_1\geq 0$ and $\lambda_3\leq 0$.
Eq.~\eqref{eq:eps_eff_eigen} shows that $\dot{\epsilon}_\mathrm{eff}$ depends not only on the strain rate of the imposed flow but also on the alignment between the micellar direction $\hat{\bm{\ell}}$ and the principal directions of the strain-rate tensor.
Specifically, $\dot{\epsilon}_\mathrm{eff}$ incorporates contributions from all three eigenvalues of $\bm{D}$, each weighted by the corresponding orientational factor $\langle\cos^2\theta_i\rangle$.
For example, $\dot{\epsilon}_\mathrm{eff}$ becomes large when the largest eigenvalue of the strain-rate tensor is large and micelles align with the corresponding eigenvector.
In contrast, $\dot{\epsilon}_\mathrm{eff}$ can be small even in very strong flows if micelles preferentially align with the eigenvector corresponding to the smallest eigenvalue.

\begin{figure}
  \centering
  \begin{overpic}[width=0.9\linewidth]{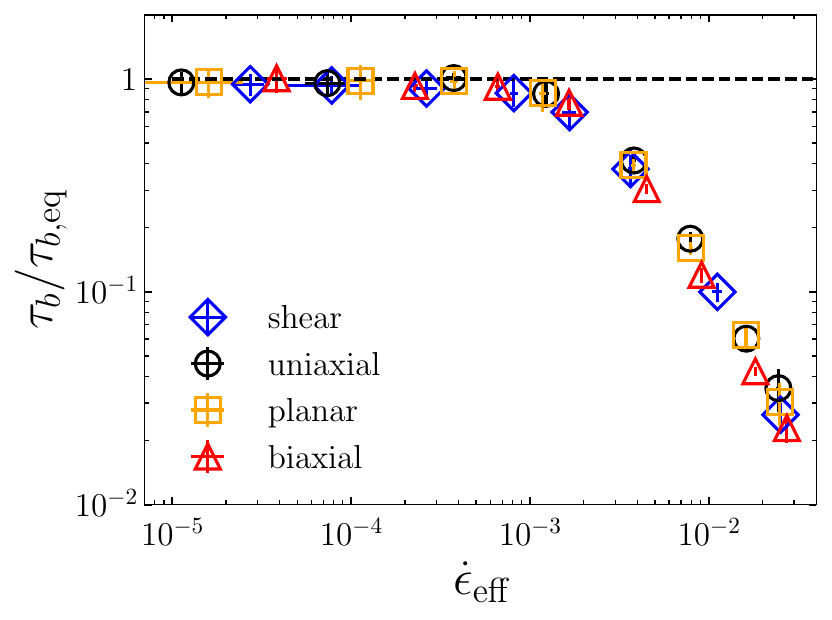} 
  \end{overpic}
  \caption{Normalized average lifetime $\tau_b/\tau_{b,\mathrm{eq}}$ of micelles with $N_\mathrm{ag}=300$ at $k_\mathrm{B}T=1$ as a function of the effective extension rate $\dot{\epsilon}_\mathrm{eff}$ for shear flow, uniaxial extensional flow, planar extensional flow, and biaxial extensional flow. The black dashed line indicates $\tau_b/\tau_{b,\mathrm{eq}}=1$. The error bars denote the standard deviations from six independent simulations.}
  \label{fig:lifetime_eps_eff}
\end{figure}

To demonstrate the relevance of $\dot{\epsilon}_\mathrm{eff}$ for flow-induced scission of wormlike micelles, we show $\tau_b/\tau_{b,\mathrm{eq}}$ for different flow types as a function of $\dot{\epsilon}_\mathrm{eff}$ in Fig.~\ref{fig:lifetime_eps_eff}.
We find that $\tau_b/\tau_{b,\mathrm{eq}}$ collapses onto a single curve irrespective of the flow type. 
The discrepancy between shear and extensional flows observed in Fig.~\ref{fig:lifetime_flow_type} is nearly removed by considering the alignment of wormlike micelles with respect to the extensional direction via Eq.~\eqref{eq:eps_eff}.
Moreover, the slightly smaller value of $\tau_b/\tau_{b,\mathrm{eq}}$ in biaxial extensional flow is accounted for by a slight increase in $\dot{\epsilon}_\mathrm{eff}$.
We therefore demonstrate that the flow-kinematics dependence of flow-induced scission can be captured by the effective extension rate $\dot{\epsilon}_\mathrm{eff}$.
We note that the definition of $\dot{\epsilon}_\mathrm{eff}$ applies to an arbitrary velocity gradient tensor $\nabla\bm{u}$ [Eq.~\eqref{eq:eps_eff}], including flows with mixed shear and extensional kinematics.~\cite{Fuller1980-ag}
The effectiveness of $\dot{\epsilon}_\mathrm{eff}$ in describing flow-induced scission in such mixed flows remains to be examined.

\begin{figure}
  \centering
  \begin{overpic}[width=0.9\linewidth]{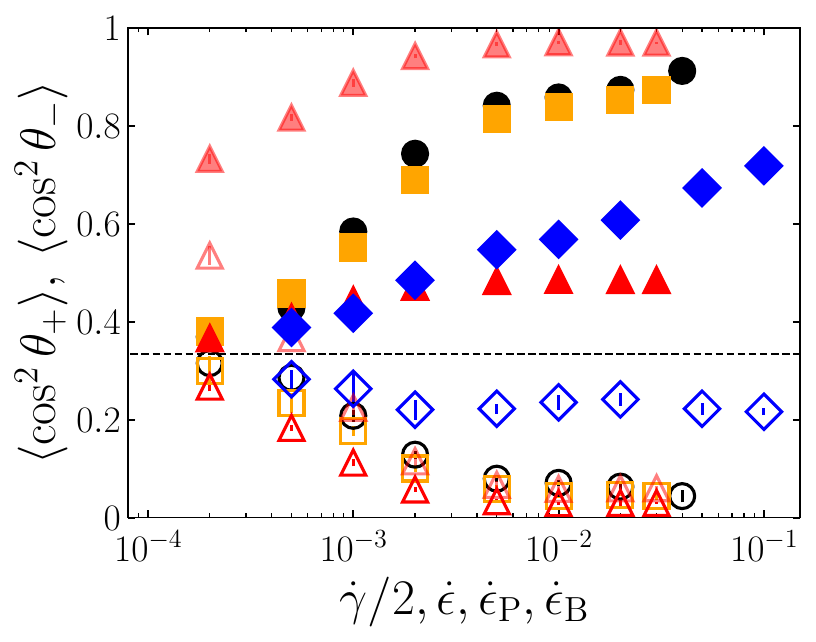} 
  \end{overpic}
  \caption{Mean-square cosines $\langle\cos^2\theta_+\rangle$~(filled symbols) and $\langle\cos^2\theta_-\rangle$~(open symbols) of the angles $\theta_+$ and $\theta_-$ between the micellar direction and the eigenvectors of the strain-rate tensor corresponding to the extensional and compressional directions, respectively, as functions of $\dot{\gamma}/2$~(blue diamond), $\dot{\epsilon}$~(black circle), $\dot{\epsilon}_\mathrm{P}$~(orange square), and $\dot{\epsilon}_\mathrm{B}$~(red triangle) for $N_\mathrm{ag}=300$ and $k_\mathrm{B}T=1$. The semitransparent red triangles show $2\langle\cos^2\theta_+\rangle$ and $2\langle\cos^2\theta_-\rangle$ for biaxial extensional flow. The black dashed line indicates $\langle\cos^2\theta_\pm\rangle=1/3$, which corresponds to the isotropic case. The error bars denote the standard deviations from six independent simulations.}

  \label{fig:alignment_flow_type}
\end{figure}

To understand the physical mechanism underlying this collapse of $\tau_b/\tau_{b,\mathrm{eq}}$ in terms of $\dot{\epsilon}_\mathrm{eff}$, we focus on the alignment behavior of wormlike micelles in different flows.
Using Eq.~\eqref{eq:eps_eff_eigen}, $\dot{\epsilon}_\mathrm{eff}$ can be written as 
\begin{equation}
  \dot{\epsilon}_\mathrm{eff}=
\begin{cases}
\dot{\gamma}/2\left(\langle\cos^2\theta_+\rangle-\langle\cos^2\theta_-\rangle\right) & (\text{shear}),\\
\dot{\epsilon}\left(\langle\cos^2\theta_+\rangle-\langle\cos^2\theta_-\rangle\right) & (\text{uniaxial}),\\
\dot{\epsilon}_\mathrm{P}\left(\langle\cos^2\theta_+\rangle-\langle\cos^2\theta_-\rangle\right) & (\text{planar}),\\
\dot{\epsilon}_\mathrm{B}\left(2\langle\cos^2\theta_+\rangle-2\langle\cos^2\theta_-\rangle\right) & (\text{biaxial}),
\end{cases}
\label{eq:eps_eff_eigen_type}
\end{equation}
where $\theta_+$ and $\theta_-$ are the angles between $\hat{\bm{\ell}}$ and the eigenvectors corresponding to extensional and compressional directions, respectively.
Figure~\ref{fig:alignment_flow_type} shows $\langle\cos^2\theta_+\rangle$ and $\langle\cos^2\theta_-\rangle$ as functions of $\dot{\gamma}/2$, $\dot{\epsilon}$, $\dot{\epsilon}_\mathrm{P}$, and $\dot{\epsilon}_\mathrm{B}$.
We observe that $\langle\cos^2\theta_+\rangle$ is smaller in shear flow than in uniaxial and planar extensional flows, indicating that micelles are less aligned with the extensional direction in shear flow.
This is because micelles tend to align in the flow direction with increasing $\dot{\gamma}$~\cite{Koide2023-ao}, while the extensional direction is at an angle of $\pi/4$ to the flow direction.
Accordingly, flow-induced scission is promoted less under shear flow~(Fig.~\ref{fig:lifetime_flow_type}).
Note that although $\langle\cos^2\theta_+\rangle$ takes the smallest value in biaxial extensional flow, biaxial extensional flow has two extensional directions, which gives rise to the prefactor 2 before $\langle\cos^2\theta_+\rangle$[see Eq.~\eqref{eq:eps_eff_eigen_type}].
Overall, the total contribution of micellar alignment along the extensional directions is slightly larger in biaxial extensional flow at large deformation rates~(see the semitransparent triangles in Fig.~\ref{fig:alignment_flow_type}).
Thus, biaxial extensional flow is somewhat more effective for flow-induced scission than uniaxial and planar extensional flows~(Fig.~\ref{fig:lifetime_flow_type}).
This result may be particularly relevant to wormlike micelles in turbulent flows, where local strain fields are known to be statistically biased toward biaxial extension~\cite{Lund1994-uv,Meneveau2011-gm}.
Such biaxial-extensional characteristics are also considered important for polymer stretching in turbulence~\cite{Zhou2003-gk,Terrapon2004-xb}.

The present definition of $\dot{\epsilon}_\mathrm{eff}$ is based on the global micellar orientation vector $\hat{\bm{\ell}}$, which is sufficient for incorporating the effect of micellar alignment in the present system.
For much longer wormlike micelles, however, local micellar orientation may become important because micellar scission occurs locally.
An important direction for future work is to refine the definition of $\dot{\epsilon}_\mathrm{eff}$ by using local direction vectors along wormlike micelles.
It is also crucial to develop a theoretical framework that enables an a priori prediction of $\dot{\epsilon}_\mathrm{eff}$, because the present formulation relies on the micellar orientation $\hat{\bm{\ell}}$ obtained from DPD simulations under flow.
Developing such a framework requires a deeper understanding of the coupling between micellar alignment and scission under flow, as discussed below.

Although $\dot{\epsilon}_\mathrm{eff}$ successfully captures the flow-kinematics dependence of flow-induced scission at fixed $k_\mathrm{B}T$, the physically relevant parameter is the dimensionless product of $\dot{\epsilon}_\mathrm{eff}$ and a characteristic timescale of wormlike micelles, rather than $\dot{\epsilon}_\mathrm{eff}$ itself.
Therefore, we aim to identify this characteristic timescale by investigating wormlike micellar solutions at different temperatures.
Because temperature affects not only the dynamics of wormlike micelles but also their scission and recombination kinetics~\cite{Koide2022-bp,Koide2023-yb}, a systematic investigation at different $k_\mathrm{B}T$ also serves to assess the applicability of $\dot{\epsilon}_\mathrm{eff}$ to describing flow-induced scission.
Here, we introduce the longest relaxation time $\tau_\Lambda$ of wormlike micelles.
In our previous study~\cite{Koide2022-bp}, we proposed a method for estimating $\tau_\Lambda$ from the micellar rotational relaxation time $\tau_r$ and the average micellar lifetime $\tau_b$.
Here, $\tau_r$ is determined from the exponential decay of the autocorrelation function of $\hat{\bm{\ell}}$.
We define $\tau_\Lambda$ as the relaxation time at the intersection of the $N_\mathrm{ag}$-dependent curves of $\tau_r$ and $\tau_b$ evaluated at equilibrium.
Because $\tau_r$ increases monotonically with $N_\mathrm{ag}$, whereas $\tau_b$ decreases monotonically, $\tau_\Lambda$ can be regarded as an estimate of the upper limit of the micellar relaxation timescale that can be realized in the system.
For micelles with $\tau_r\gtrsim \tau_\Lambda$, scission typically occurs before rotational relaxation is completed because their average lifetime satisfies $\tau_b\lesssim \tau_r$.
Consequently, relaxation modes with timescales longer than $\tau_\Lambda$ are effectively suppressed by micellar scission.
The relevance of $\tau_\Lambda$ has been demonstrated for micellar alignment under shear flow~\cite{Koide2023-ao} and the steady-state extensional viscosity~\cite{Koide2025-zr}.
In the present study, we consider three temperatures $k_\mathrm{B}T=0.9,\,1,\,1.2$.
The corresponding values of $\tau_\Lambda$ are $1450$, $767$, and $271$, respectively.
Using $\tau_\Lambda$, we define the Weissenberg numbers for the different flow types as $\mathrm{Wi}=\tau_\Lambda \dot{\gamma}$, $\mathrm{Wi}_\mathrm{E}=\tau_\Lambda \dot{\epsilon}$, $\mathrm{Wi}_\mathrm{P}=\tau_\Lambda \dot{\epsilon}_\mathrm{P}$, and $\mathrm{Wi}_\mathrm{B}=\tau_\Lambda \dot{\epsilon}_\mathrm{B}$.
We also define the effective Weissenberg number as $\mathrm{Wi}_\mathrm{eff}=\tau_\Lambda\dot{\epsilon}_\mathrm{eff}$.

\begin{figure}
    \centering
        \begin{tabular}{c}
        \begin{minipage}{1\hsize}
    \centering
            \begin{overpic}[width=0.72\linewidth]{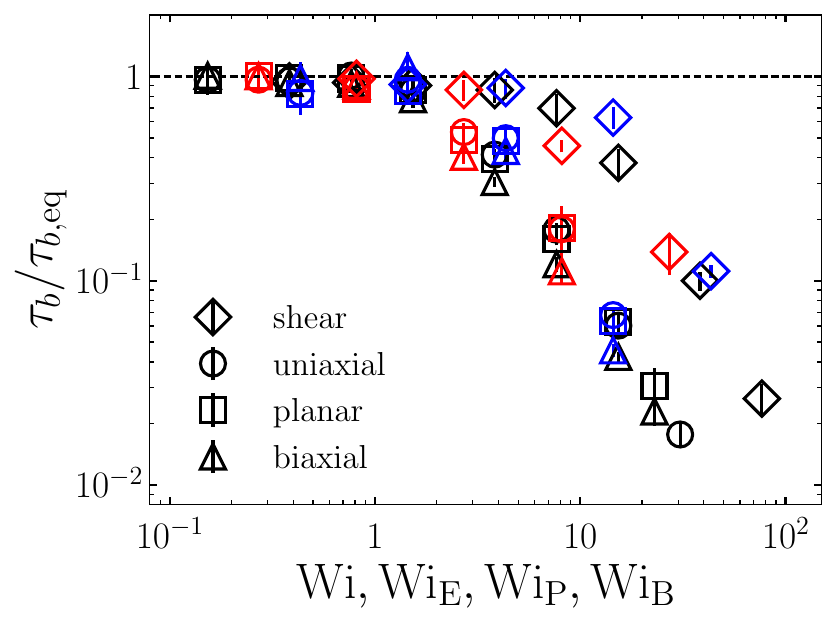}
                \linethickness{3pt}
            \put(5,70){(a)}
            \end{overpic}
        \end{minipage}\\
        \begin{minipage}{1\hsize}
    \centering
            \begin{overpic}[width=0.72\linewidth]{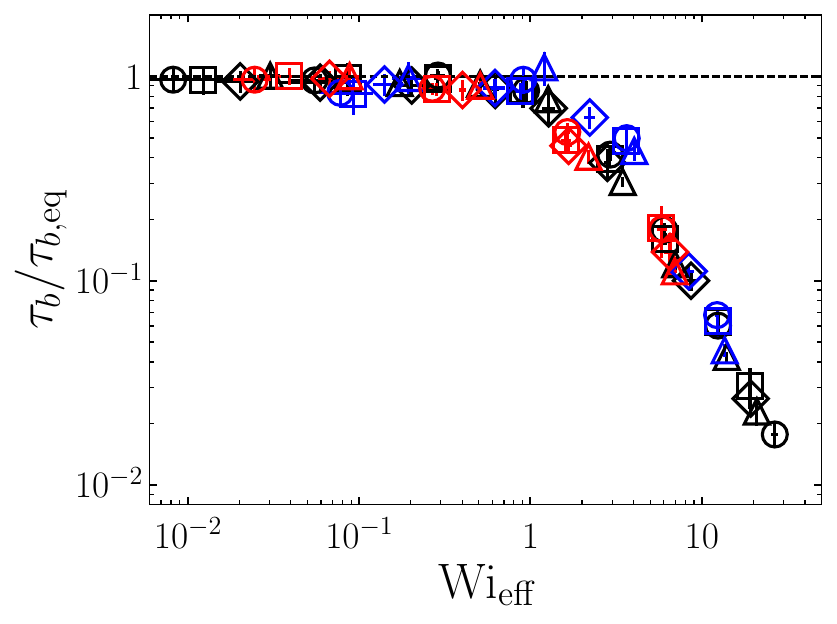}
                \linethickness{3pt}
            \put(5,70){(b)}

            \end{overpic}
        \end{minipage}
        \end{tabular}
  \caption{Normalized average lifetime $\tau_b/\tau_{b,\mathrm{eq}}$ of micelles with $N_\mathrm{ag}=300$ as a function of (a) the Weissenberg numbers $\mathrm{Wi}$, $\mathrm{Wi}_\mathrm{E}$, $\mathrm{Wi}_\mathrm{P}$, and $\mathrm{Wi}_\mathrm{B}$ and (b) the effective Weissenberg number $\mathrm{Wi}_\mathrm{eff}$. 
  Different colors correspond to different values of $k_\mathrm{B}T$: blue, $k_\mathrm{B}T=0.9$; black, $1$; red, $1.2$.
  The black dashed lines indicate $\tau_b/\tau_{b,\mathrm{eq}}=1$. 
  The error bars denote the standard deviations from six independent simulations.}
  
        \label{fig:lifetime_temperature}
  \end{figure}

We investigate the temperature dependence of the average lifetime to assess whether $\tau_\Lambda$ provides an appropriate timescale for flow-induced scission.
Figure~\ref{fig:lifetime_temperature}(a) shows $\tau_b/\tau_{b,\mathrm{eq}}$ of micelles with $N_\mathrm{ag}=300$ as a function of $\mathrm{Wi}$, $\mathrm{Wi}_\mathrm{E}$, $\mathrm{Wi}_\mathrm{P}$, and $\mathrm{Wi}_\mathrm{B}$.
As observed for $k_\mathrm{B}T=1$~(Fig.~\ref{fig:lifetime_flow_type}), flow-kinematics dependence is evident irrespective of $k_\mathrm{B}T$, i.e., $\tau_b/\tau_{b,\mathrm{eq}}$ is smaller for extensional flows than for shear flow.
For a given flow type, normalization by $\tau_\Lambda$ leads to a partial collapse of the results obtained at different temperatures; however, the collapse remains imperfect, especially for shear flows, as already reported in our previous study~\cite{Koide2022-bp}.
This limitation is overcome by using the effective Weissenberg number $\mathrm{Wi}_\mathrm{eff}$.
Indeed, Fig.~\ref{fig:lifetime_temperature}(b) shows $\tau_b/\tau_{b,\mathrm{eq}}$ as a function of $\mathrm{Wi}_\mathrm{eff}$, where all the results collapse onto a single curve irrespective of $k_\mathrm{B}T$ and flow type.
Note that the imperfect collapse for shear flow~[Fig.~\ref{fig:lifetime_temperature}(a)] is mitigated by using $\mathrm{Wi}_\mathrm{eff}$.
This tendency indicates that even for the same flow type at a fixed Weissenberg number based on the imposed deformation rate, variations in micellar alignment must be taken into account to describe the temperature dependence of flow-induced scission.

\begin{figure}
    \centering
        \begin{tabular}{c}
        \begin{minipage}{1\hsize}
    \centering
            \begin{overpic}[width=0.75\linewidth]{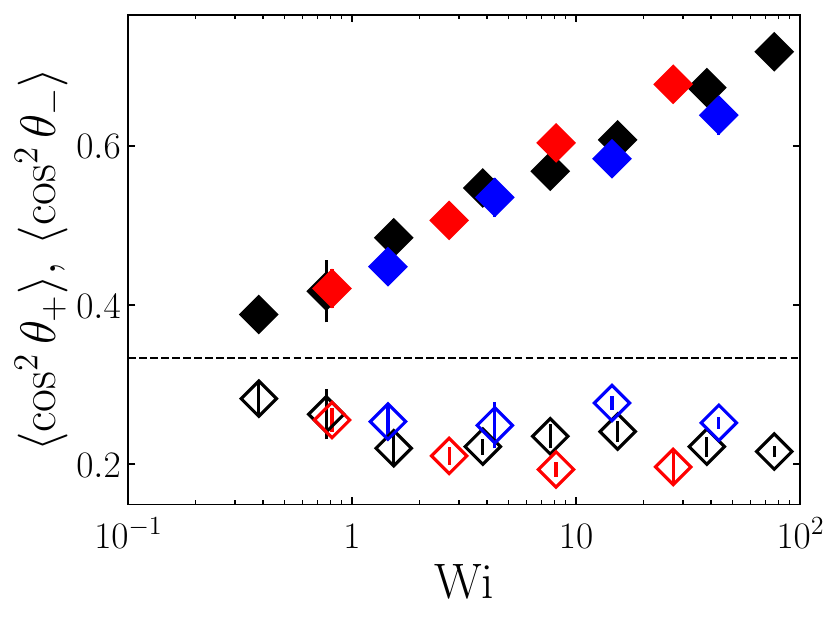}
                \linethickness{3pt}
            \put(3,70){(a)}
            \end{overpic}
        \end{minipage}\\
        \begin{minipage}{1\hsize}
    \centering
            \begin{overpic}[width=0.75\linewidth]{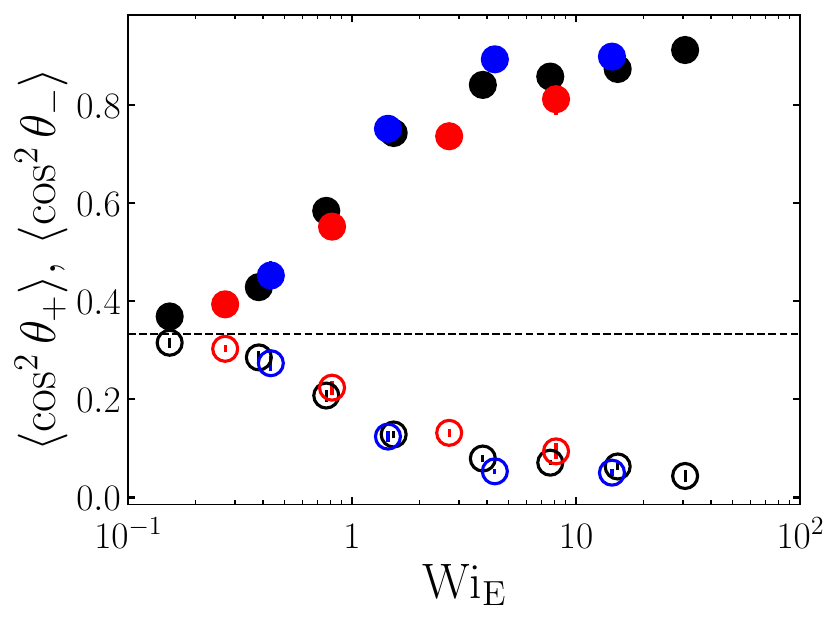}
                \linethickness{3pt}
            \put(3,70){(b)}
            \end{overpic}
        \end{minipage}\\
                \begin{minipage}{1\hsize}
    \centering
            \begin{overpic}[width=0.75\linewidth]{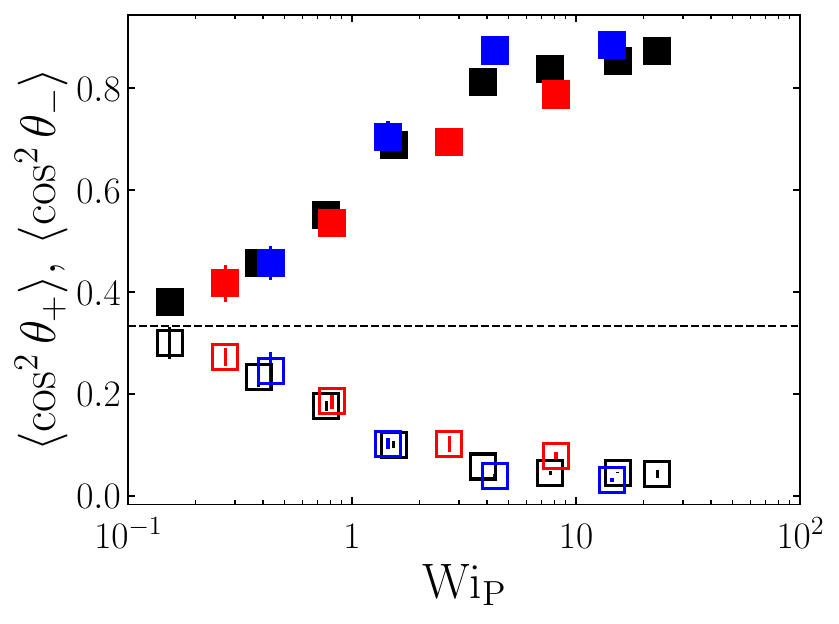}
                \linethickness{3pt}
            \put(3,70){(c)}
            \end{overpic}
        \end{minipage}\\
        \begin{minipage}{1\hsize}
    \centering
            \begin{overpic}[width=0.75\linewidth]{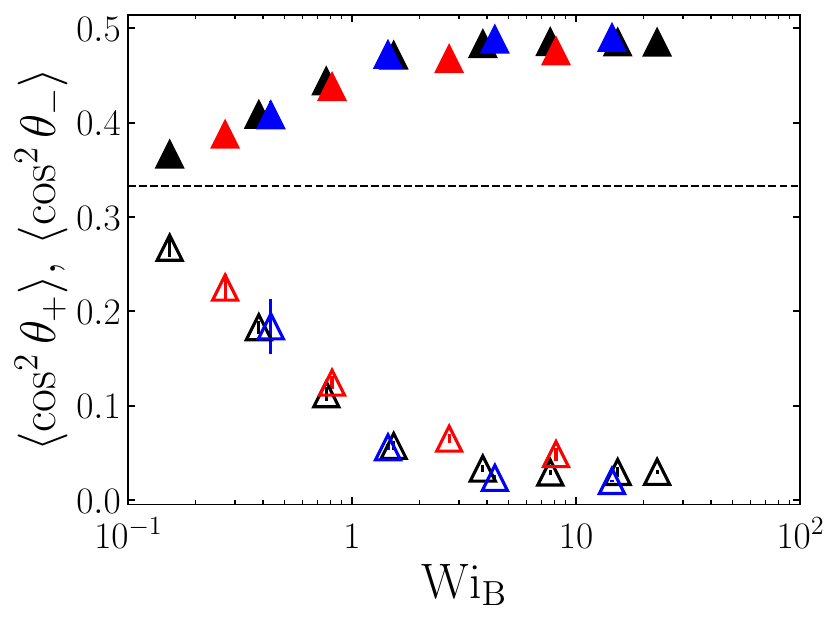}
                \linethickness{3pt}
            \put(3,70){(d)}

            \end{overpic}
        \end{minipage}
        \end{tabular}
          \caption{Mean-square cosines $\langle\cos^2\theta_+\rangle$~(filled symbols) and $\langle\cos^2\theta_-\rangle$~(open symbols) of the angles $\theta_+$ and $\theta_-$ between the micellar direction and the eigenvectors of the strain-rate tensor corresponding to the extensional and compressional directions, respectively, for $N_\mathrm{ag}=300$ at $k_\mathrm{B}T=0.9$~(blue), $1$~(black), and $1.2$~(red) in (a) shear flow, (b) uniaxial extensional flow, (c) planar extensional flow, and (d) biaxial extensional flow. The black dashed lines indicate $\langle\cos^2\theta_\pm\rangle=1/3$, which corresponds to the isotropic case. The error bars denote the standard deviations from six independent simulations.}
        \label{fig:alignment_temp}
  \end{figure}

We therefore examine the temperature dependence of micellar alignment under different flow types.
Figure~\ref{fig:alignment_temp} shows $\langle\cos^2\theta_+\rangle$ and $\langle\cos^2\theta_-\rangle$ as functions of the Weissenberg number for each flow type.
Overall, $\langle\cos^2\theta_+\rangle$ and $\langle\cos^2\theta_-\rangle$ show little dependence on $k_\mathrm{B}T$ at fixed Weissenberg numbers, particularly at low Weissenberg numbers.
In shear flow, however, $\langle\cos^2\theta_+\rangle$ ($\langle\cos^2\theta_-\rangle$) becomes slightly larger (smaller) with increasing $k_\mathrm{B}T$ at large $\mathrm{Wi}$.
This tendency can be attributed to the $k_\mathrm{B}T$ dependence of the complicated coupling between orientation dynamics and scission kinetics under shear flow~\cite{Koide2023-ao}.
This temperature-dependent alignment leads to the imperfect collapse of the shear-flow data in Fig.~\ref{fig:lifetime_temperature}(a): $\tau_b/\tau_{b,\mathrm{eq}}$ for shear flow is shifted downward for $k_\mathrm{B}T=1.2$ and upward for $k_\mathrm{B}T=0.9$.
We thus demonstrate that the effective Weissenberg number $\mathrm{Wi}_\mathrm{eff}$ is an essential dimensionless parameter for describing flow-induced scission of wormlike micelles, at least within the present model.
The importance of $\mathrm{Wi}_\mathrm{eff}$ indicates that the relative change in the degree of flow-induced scission is determined by the competition between the flow-induced stretching rate along micelles, which depends on both flow kinematics and micellar alignment, and the relaxation rate of wormlike micelles.

\section{Conclusions}
To provide a unified framework for describing flow-induced scission of wormlike micelles, we examined the deformation-rate dependence of micellar lifetime under shear, uniaxial extensional, planar extensional, and biaxial extensional flows using DPD simulations.
Regardless of the flow type, flow-induced scission of wormlike micelles occurred under sufficiently strong flows, as confirmed by the aggregation-number distribution $P(N_\mathrm{ag})$~(Fig.~\ref{fig:nag_pdf}) and the survival function $S(t)$ of micellar lifetime~(Fig.~\ref{fig:survival_function}).
Quantitatively, however, the average lifetime $\tau_b$ of wormlike micelles exhibited flow-type dependence when plotted as a function of $\dot{\gamma}$, $\dot{\epsilon}$, $\dot{\epsilon}_\mathrm{P}$, and $\dot{\epsilon}_\mathrm{B}$~(Fig.~\ref{fig:lifetime_flow_type}).
In particular, shear flow is less effective for flow-induced scission than extensional flows, and this difference cannot be fully explained by kinematic measures derived solely from the imposed velocity gradient tensor $\nabla\bm{u}$.

To rationalize this flow-type dependence, we introduced the effective extension rate $\dot{\epsilon}_\mathrm{eff}$, which incorporates not only the flow kinematics but also the alignment of wormlike micelles~[Eq.~\eqref{eq:eps_eff}].
We demonstrated that the data for $\tau_b$ obtained under different flow types collapsed onto a single curve when plotted against $\dot{\epsilon}_\mathrm{eff}$ ~(Fig.~\ref{fig:lifetime_eps_eff}).
Since $\dot{\epsilon}_\mathrm{eff}$ can be expressed in terms of the eigenvalues of the strain-rate tensor $\bm{D}$ and the angles between the corresponding eigenvectors and the micellar direction, we analyzed the alignment of wormlike micelles with the extensional and compressional directions.
Wormlike micelles aligned preferentially with the extensional directions under uniaxial, planar, and biaxial extensional flows, whereas, under shear flow, they tended to align with the flow direction, which is at an angle of $\pi/4$ to the extensional direction~(Fig.~\ref{fig:alignment_flow_type}).
Thus, shear flow promotes flow-induced scission less than extensional flows.

Furthermore, we introduced the effective Weissenberg number $\mathrm{Wi}_\mathrm{eff}=\tau_\Lambda\dot{\epsilon}_\mathrm{eff}$, defined as the product of $\dot{\epsilon}_\mathrm{eff}$ and the longest relaxation time $\tau_\Lambda$ of wormlike micelles.
This quantity serves as a key dimensionless parameter for describing flow-induced scission at different temperatures.
When plotted against the Weissenberg number defined simply as the product of $\tau_\Lambda$ and the imposed deformation rate, the normalized average lifetime $\tau_b/\tau_{b,\mathrm{eq}}$ depended on the flow type and also showed a slight temperature dependence for each flow type~[Fig.~\ref{fig:lifetime_temperature}(a)].
In contrast, the data for $\tau_b/\tau_{b,\mathrm{eq}}$ collapsed onto a single curve when plotted against $\mathrm{Wi}_\mathrm{eff}$~[Fig.~\ref{fig:lifetime_temperature}(b)].
This successful collapse indicates that $\mathrm{Wi}_\mathrm{eff}$ captures both differences among flow types and subtle temperature-dependent variations in micellar alignment within each flow type~(Fig.~\ref{fig:alignment_temp}).
We therefore concluded that the degree of flow-induced scission is governed by the competition between the relaxation rate of wormlike micelles and their flow-induced stretching rate, which is determined by both flow kinematics and micellar alignment.

Note, however, that our results are restricted to a single surfactant model.
Since the energetic cost of micellar scission may vary among surfactant types, the master curve shown in Fig.~\ref{fig:lifetime_temperature}(b) may also depend on the surfactant type.
Although the physical origin of its functional form could not be identified in the present study, systematic comparisons among different surfactant types may provide insight into the underlying mechanism. 
This remains an important subject for future study.
\section*{Author contributions}
Yusuke Koide: Conceptualization, Data curation, Formal analysis, Funding acquisition, Investigation, Methodology, Project administration, Software, 
 Validation, Visualization, Writing -- original draft, Writing -- review \& editing.
Takato Ishida: Investigation, Writing -- original draft, Writing -- review \& editing.
Takashi Uneyama: Investigation, Writing -- original draft, Writing -- review \& editing.
Yuichi Masubuchi: Investigation, Writing -- original draft, Writing -- review \& editing.
\section*{Conflicts of interest}
There are no conflicts to declare.

\section*{Data availability}
The data supporting the findings of this study are included in the article.
Additional raw data are available from the corresponding author upon reasonable request.

\section*{Appendix A: Aggregation-number dependence of flow-induced scission}
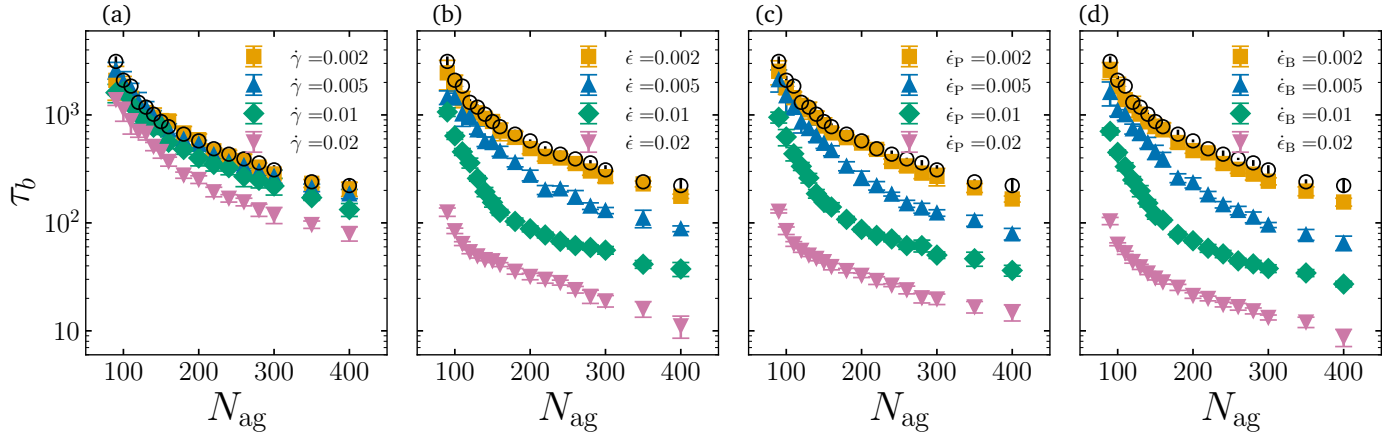
\begin{figure*}
  \centering
  \begin{overpic}[width=1\linewidth]{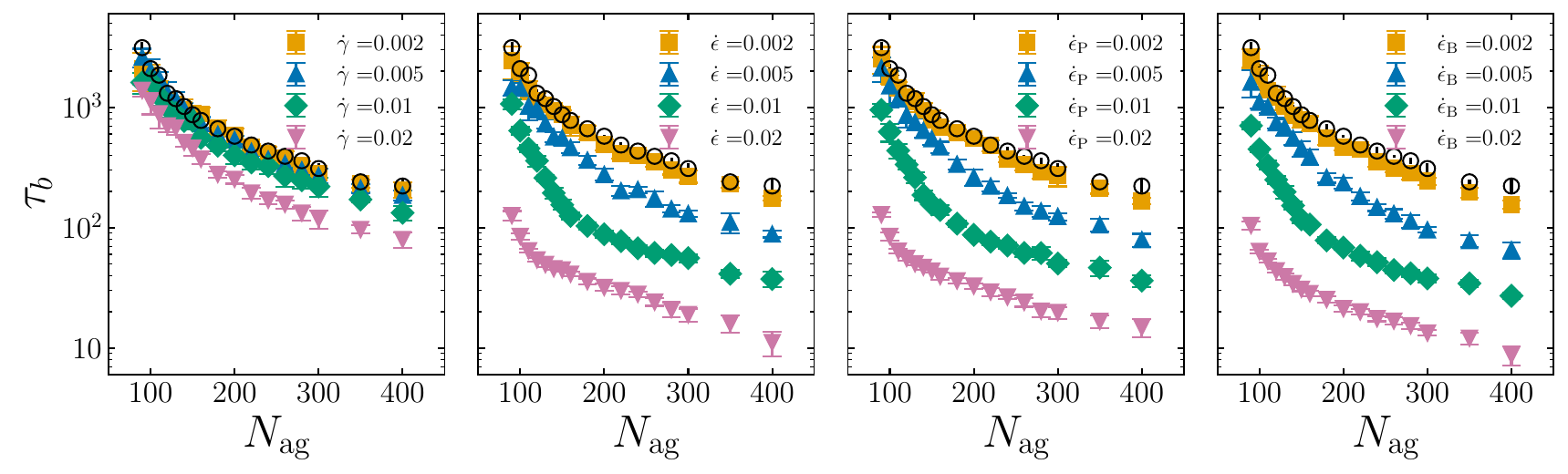} 
            \put(8,30){(a)}
            \put(31.5,30){(b)}
            \put(54.5,30){(c)}
            \put(77.5,30){(d)}
  \end{overpic}
  \caption{Average lifetime $\tau_b$ of micelles as a function of the aggregation number $N_\mathrm{ag}$ at $k_\mathrm{B}T=1$ under (a) shear flow, (b) uniaxial extensional flow, (c) planar extensional flow, and (d) biaxial extensional flow. The black open circles represent the results at equilibrium.}

  \label{fig:lifetime_nag}
\end{figure*}

In the main text, we focused on micelles with aggregation numbers of $N_\mathrm{ag}= 300$.
Here, we examine a broader range of $N_\mathrm{ag}$ to provide supplementary information on the $N_\mathrm{ag}$ dependence of flow-induced scission and to confirm that our conclusions remain essentially unchanged in the range of $N_\mathrm{ag}$ where wormlike micelles are formed.
Figure~\ref{fig:lifetime_nag} shows the average lifetime $\tau_{b}$ of micelles as a function of $N_\mathrm{ag}$ for each flow type.
At high deformation rates, $\tau_b$ becomes smaller than its equilibrium value irrespective of $N_\mathrm{ag}$, indicating the occurrence of flow-induced scission.
For $N_\mathrm{ag}\gtrsim 200$, increasing the deformation rate shifts $\tau_b$ downward nearly in parallel on the semilogarithmic plot, indicating that $\tau_b$ is reduced by an approximately constant factor independent of $N_\mathrm{ag}$.
For $N_\mathrm{ag}\lesssim 200$, however, this uniform reduction is no longer observed, and the relative decrease in $\tau_b$ depends on $N_\mathrm{ag}$.

We attribute this behavior to two main factors.
The first is the $N_\mathrm{ag}$ dependence of the micellar relaxation time.
As shown in previous studies~\cite{Koide2022-bp,Koide2023-ao,Koide2025-zr}, there exists a characteristic aggregation number \(N_\Lambda\).
For $N_\mathrm{ag}\lesssim N_\Lambda$, the micellar relaxation time increases with $N_\mathrm{ag}$, whereas for $N_\mathrm{ag}\gtrsim N_\Lambda$, micelles are characterized by a common relaxation time $\tau_\Lambda$.
Since $N_\Lambda\simeq 163$ in the present system at $k_\mathrm{B}T=1$, the $N_\mathrm{ag}$ dependence of $\tau_{b}$, particularly around $N_\mathrm{ag}\simeq 200$, can be attributed to the dependence of the relaxation time on $N_\mathrm{ag}$.
The second possible factor is the morphology dependence of flow-induced scission.
In the systems considered, micelles are rodlike for $50 \lesssim N_\mathrm{ag}\lesssim 200$ and wormlike for $N_\mathrm{ag}\gtrsim 200$~\cite{Koide2022-bp}.
Thus, changes in micellar morphology around $N_\mathrm{ag}\simeq 200$ may also contribute to the observed $N_\mathrm{ag}$ dependence of $\tau_{b}$.
Since $N_\Lambda$ depends on temperature and the interaction parameters, a more systematic investigation over a wider range of conditions could separate the effects of relaxation time and micellar morphology on $\tau_b$.
Such an analysis, however, is beyond the scope of the present study.

\begin{figure}
  \centering
  \begin{overpic}[width=0.9\linewidth]{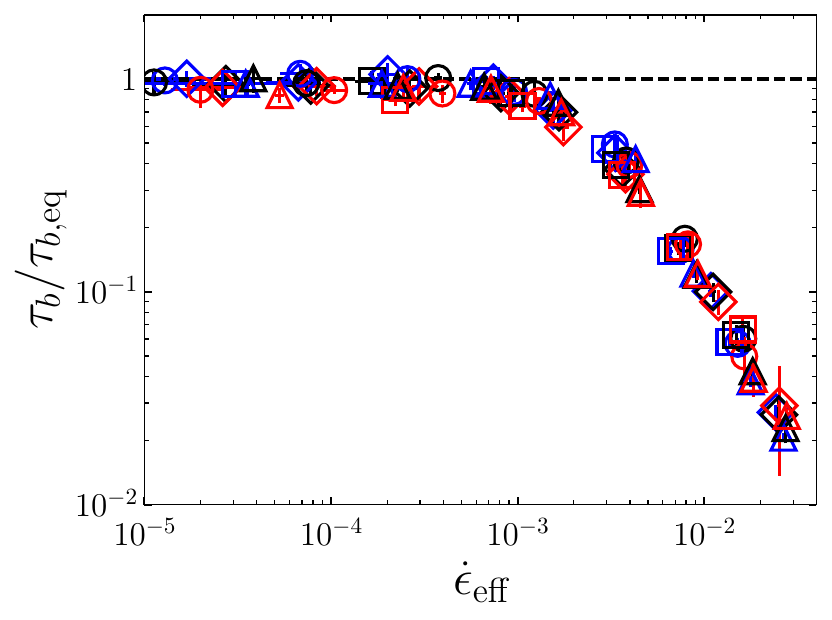} 
  \end{overpic}
  \caption{Normalized average lifetime $\tau_b/\tau_{b,\mathrm{eq}}$ of micelles with $N_\mathrm{ag}=200$~(blue), $300$~(black), and $400$~(red) as a function of the effective extension rate $\dot{\epsilon}_\mathrm{eff}$ at $k_\mathrm{B}T=1$ under shear flow~(diamond), uniaxial extensional flow~(circle), planar extensional flow~(square), and biaxial extensional flow~(triangle). The black dashed line indicates $\tau_b/\tau_{b,\mathrm{eq}}=1$. The error bars denote the standard deviations from six independent simulations.}
  \label{fig:lifetime_eps_eff_nag}
\end{figure}

Since the present study is concerned primarily with wormlike micelles, we focus on the range $N_\mathrm{ag}\gtrsim 200$.
Figure~\ref{fig:lifetime_eps_eff_nag} shows $\tau_{b}/\tau_{b,\mathrm{eq}}$ as a function of the effective extension rate $\dot{\epsilon}_\mathrm{eff}$ for $N_\mathrm{ag}=200$, $300$, and $400$ under shear, uniaxial extensional, planar extensional, and biaxial extensional flows.
The collapse of the results for different $N_\mathrm{ag}\gtrsim 200$ onto a single curve further supports our conclusion that the effective extension rate $\dot{\epsilon}_\mathrm{eff}$~[Eq.~\eqref{eq:eps_eff}] is a key quantity governing flow-induced scission of wormlike micelles.
\section*{Appendix B: Derivation of the effective extension rate for material lines}
To clarify the physical meaning of the effective extension rate, we consider a material line represented by two nodes $\bm{x}_1(t)$ and $\bm{x}_2(t)$, which are advected by the local velocity as 
\begin{equation}
  \frac{d\bm{x}_i(t)}{dt} = \bm{u}(\bm{x}_i(t),t),
\end{equation}
where $\bm{u}(\bm{x},t)$ is the imposed velocity field.
The vector connecting the two nodes $\bm{\ell}(t)=\bm{x}_2(t)-\bm{x}_1(t)$ evolves according to
\begin{equation}
  \frac{d\bm{\ell}(t)}{dt} = \bm{u}(\bm{x}_2(t),t)-\bm{u}(\bm{x}_1(t),t).\label{eq:line}
\end{equation}
This simple model captures the elementary process of micellar stretching due to the velocity difference between the two ends.
In the present study, we consider flows with a time-independent and spatially uniform velocity gradient $\nabla\bm{u}$.
Thus, Eq.~\eqref{eq:line} can be written as 
\begin{equation}
  \frac{d\bm{\ell}(t)}{dt} = (\nabla \bm{u})^\mathsf{T}\cdot \bm{\ell}(t).
\end{equation}
Taking the inner product of both sides with $\hat{\bm{\ell}}(t)=\bm{\ell}(t)/|\bm{\ell}(t)|$, we obtain the evolution equation for the line length $\ell(t)$:
\begin{equation}
  \frac{d{\ell}(t)}{dt} = \left[\hat{\bm{\ell}}(t)\cdot \bm{D}\cdot \hat{\bm{\ell}}(t)\right]\ell(t),\label{eq:line_length}
\end{equation}
where $\bm{D}=\{\nabla\bm{u}+(\nabla\bm{u})^\mathsf{T}\}/2$ is the strain-rate tensor, corresponding to the symmetric part of $\nabla\bm{u}$. 
Eq.~\eqref{eq:line_length} naturally leads to the definition of the effective extension rate as
\begin{equation}
  \dot{\epsilon}_\mathrm{eff}=\langle\hat{\bm{\ell}}\cdot \bm{D}\cdot \hat{\bm{\ell}}\rangle.
\end{equation}
Thus, $\dot{\epsilon}_\mathrm{eff}$ represents the average rate at which the material line is stretched or compressed along its own direction.
\section*{Acknowledgements}
This work was supported by JSPS Grants-in-Aid for Scientific Research (24KJ0109) and JST ACT-X (JPMJAX24D5). 
The DPD simulations were mainly conducted under the HPCI Research Projects using the supercomputer ``Flow'' at Information Technology Center, Nagoya University~(hp250196).




\balance


\bibliographystyle{rsc} 
\providecommand*{\mcitethebibliography}{\thebibliography}
\csname @ifundefined\endcsname{endmcitethebibliography}
{\let\endmcitethebibliography\endthebibliography}{}

\end{document}